\documentclass[fleqn,usenatbib]{mnras}
\usepackage{newtxtext,newtxmath}
\usepackage[T1]{fontenc}

\usepackage{graphicx}	% Including figure files
\usepackage{amsmath}	% Advanced maths commands
\usepackage{color}

\def\wisk#1{\ifmmode{#1}\else{$#1$}\fi}
\def\wm2sr {Wm$^{-2}$sr$^{-1}$ }		%new definition
\def\nw2m4sr2 {nW$^2$m$^{-4}$sr$^{-2}$\ }		% new definition
\def\nwm2sr {nWm$^{-2}$sr$^{-1}$\ }		% new definition
\def\nw2m4sr {nW$^2$m$^{-4}$sr$^{-1}$\ }

\title[Short title, max. 45 characters]{MNRAS \LaTeXe\ template -- title goes here}

\title[The UHZ1-type systems by primordial black holes]{A possible pathway to UHZ1-type systems at $z\sim 10$ by heterogeneous mass primordial black holes as dark matter}

\author[A. Kashlinsky, F. Atrio-Barandela \& D. Martin-Gonzalez]{
Alexander Kashlinsky$^{1}$\thanks{E-mail: Alexander.Kashlinsky@nasa.gov},
Fernando Atrio-Barandela$^{2}$\thanks{E-mail: atrio@usal.es}
and Diego Mart{\'\i}n-Gonz\'alez$^{2}$\thanks{E-mail: diegomg@usal.es}\\
$^{1}$ Code 665, Observational Cosmology Lab, NASA Goddard Space Flight Center,\\
Greenbelt, MD 20771, Dept of Astronomy, University of Maryland, College Park, MD  20742, and \\
CRESST, Center for Research and Exploration in Space Science and Technology, NASA/GSFC, Greenbelt, MD 20771, USA\\
$^{2}$ Department of Fundamental Physics, University of Salamanca, 37008 Salamanca, Spain
}

\date{Accepted XXX. Received YYY; in original form ZZZ}

\pubyear{2025}

\begin{document}
\label{firstpage}
\pagerange{\pageref{firstpage}--\pageref{lastpage}}
\maketitle

\begin{abstract}
Recent space-based observations discovered several unusual objects, exhibiting similar properties, at redshifts $z\gtrsim 10$. Among them is the UHZ1 system at $z=10.1$, containing $\sim 10^8M_\odot$ in stars, with a similarly massive central black hole of $\sim 10^{7-8}M_\odot$. Here we propose a possible mechanism for forming such systems which hinges on the presence of primordial black holes (PBHs) covering a range of masses while contributing a significant fraction of the dark matter (DM). We evaluate the accurate expression for the small-scale power responsible for the collapse of the first halos in the presence of the PBH population. The extra power in the matter density field, produced by the granulation term, will cause an earlier collapse of DM halos, populated by PBHs of different masses. In these collapsed and virialized systems the PBHs will undergo 2-body relaxation, driving the more massive PBHs to the halo center under dynamical friction. We quantify this evolution for a distribution of PBH orbital parameters and halo properties. The analysis shows that PBHs can have appropriate mass functions capable of producing systems with parameters similar to what is observed for UHZ1. We suggest that the proposed mechanism could account for a subset of other systems newly discovered with the JWST at high redshifts, including the Little Red Dots. %This argues that at least a part of the DM could be composed of PBHs. 
\end{abstract}


\maketitle

\section{Introduction} \label{sec:intro}

Recent observations by JWST and Chandra revealed unusually massive objects at redshift $z\gtrsim 10$. Of particular relevance here is the UHZ1 system with its intriguing properties inferred from the observations: 1) a stellar mass of $M_\star\sim (1.4^{+0.3}_{-0.4})\times 10^8M_\odot$ at $z=10.1$, as indicated by the spectroscopic redshift measurement by JWST \citep{Goulding:2023}; and 2) an X-ray bolometric luminosity of $L_{\rm X} \sim 5\times10^{45}$ erg/s, originating in a (presumably) central black hole (BH) with $M_{\rm BH}\sim 10^7-10^8 M_\odot$, as suggested by Chandra observations \citep{Bogdan:2024}.
The GN-z11 system at $z=10.6$ has an estimated central BH and stellar masses of $10^{6.2\pm 0.3}M_\odot$ \citep{Maiolino:2024} and $\sim 10^9M_\odot$  \citep{Oesch:2016}, respectively. 
The parameters of the GHZ9 system at redshift $z=10.15$ are still more uncertain with stellar masses in the range $M_\star\simeq 5\times 10^7-3\times 10^8M_\odot$
and $M_{\rm BH}=(8.0^{+3.7}_{-3.2})\times 10^7M_\odot$  \citep{Kovacs:2024,Napolitano2024}.
The CANUCS-LRD-z8.6 system at z=8.6 has a
central BH whose inferred mass is $M_{BH}=1.0^{+0.6}_{-0.4}\times 10^8M_\odot$
\citep{Tripodi:2025}. 
The deduced BH to stellar mass ratios for many of the newly discovered high-$z$ systems appear 2--3 orders of magnitude higher than in the present day Universe \citep{Goulding:2023}. Light ($10-10^2M_\odot$) and heavy ($10^4-10^5M_\odot$) mass
BHs have been suggested as seeds of supermassive BHs (SMBHs) observed at redshifts $z\geq 6$ \citep[e.g.,][]{Smith:2019,Woods2019,Inayoshi2020}.
UHZ1, in particular, has been argued to come from a direct collapse black hole \citep{Natarajan:2024}, which may provide a formation pathway for such over-massive systems
\citep[e.g.,][]{Habouzit2016,Chon2018,Wise:2019,Latif:2022,Jeon2025,Prole:2025}, although explaining the increasing number
of these SMBHs at high redshifts is becoming a challenge to current theoretical models of BH formation \citep{Tripodi:2025}.

Here we propose an alternative origin for such objects at $z\sim 10$, assuming that primordial black holes (PBHs) make up at least part of the dark matter (DM), as proposed originally by 
\cite{Bird:2016,Kashlinsky:2016,Clesse:2017}, following the first LIGO gravitational wave discovery of two similar mass, low spin BHs \citep{Abbott:2016b}. 
Astrophysical aspects of the possibility were recently utilized in simulations by \cite{Dayal:2025,Prole:2025,Zhang:2025}. Here we specifically assess whether dynamical friction in a collapsed halo made of PBHs with heterogeneous mass distribution can successfully drive the most massive PBHs to the halo center reproducing the observed parameters at $z\gtrsim10$. Dynamical friction operates when the halo collapses and virializes and is independent of star formation. 
As discussed in \cite{Lacki:2010} the PBH DM fraction, $f_{\rm PBH}$, is expected to be either close to unity or negligible if DM consists of WIMP type, self-annihilating particles. If $f_{\rm PBH}$ were significant, the isocurvature component of the density field would dominate the inflationary component on small scales, and the first halos would collapse and virialize significantly earlier \citep{Kashlinsky:2016,Carr:2018,Kashlinsky:2021} than in standard $\Lambda$CDM cosmology. Recent measurements of source-subtracted cosmic infrared background (CIB) anisotropies from {\it Spitzer}/IRAC \citep{Kashlinsky:2025} appear to provide strong evidence for the PBH-DM connection; see \cite{Kashlinsky:2016} and CIB review by \cite{Kashlinsky:2018}. 
We show that with a suitable PBH mass-function the most massive PBHs would sink to the halo centers to form a dominant BH of the required mass by $z\sim10$, while the baryonic gas would naturally form stars,  as observed for UHZ1.
The UHZ1 properties that we seek to explain within the framework of the PBH-DM scenario are: 1) the existence of the $\sim 10^7-10^8\,M_\odot$ (presumably central) BH, which accounts for a few percent of the collapsed halo mass, and 2) the existence of the early collapsed halo  at $z=10.1$, containing $\sim 10^8\,M_\odot$ in stars, with 3) an implied total halo mass $\gtrsim 10^9\,M_\odot$. 

Theoretical motivations for PBHs in the early Universe range from modifications of the inflationary model \citep{Garcia-Bellido:1996} to phase transitions in the early Universe \citep{Jedamzik:1997}. If so, it is generally expected that PBHs would have a broad mass spectrum, which currently is not probed directly, but theoretical models have been developed \citep{Byrnes:2018,Carr:2021a,Carr:2021,Carr:2021b}. The mass function may reflect the PBH formation epochs, such as the QCD phase transition, which would imprint masses not far from the typical LIGO-observed objects. Some mass functions would allow for significant gravitational dynamical evolution of the PBHs, with the more massive ones sinking to the halo centers due to dynamical friction to efficiently form a central object, as is explored below. The gas would then collapse inside such halos by $z\sim10$ to form stars.
We here propose that this mechanism may have led to the formation of UHZ1-type systems with the observed parameters, provided that the currently unknown PBH mass function is of a suitable type, as will be discussed in more detail below. 

We adopt standard cosmological parameters for a flat Universe with $\Omega_{\rm m}=0.28$, and $h=0.71$. The cosmic time at $z\gtrsim 10$ is $t_{\rm cosm}\simeq 0.2(1+z/20)^{-3/2}$~Gyr, with a corresponding Hubble parameter of $H(z)\simeq 71\, (1+z)^{3/2}$ km~s$^{-1}$\,Mpc$^{-1}$. In Section~\ref{sec:evolution}, we discuss the evolution of the pre-collapse density field in the presence of PBHs as DM and present an accurate analytical expression for the total density power responsible for first halo collapse in the presence of the PBHs. In Section~\ref{sec:sec3}, we consider the effects of dynamical friction and mass segregation in the subsequently collapsed halos, containing PBHs of different masses, leading to central UHZ1 mass BH formation prior to $z\sim10$, and briefly comment on the behavior of the baryonic (gaseous) component. We conclude with a brief discussion.

\section{Evolution of density field within PBH-DM cosmology}
\label{sec:evolution}

\begin{figure*}
\includegraphics[width=6.25in]{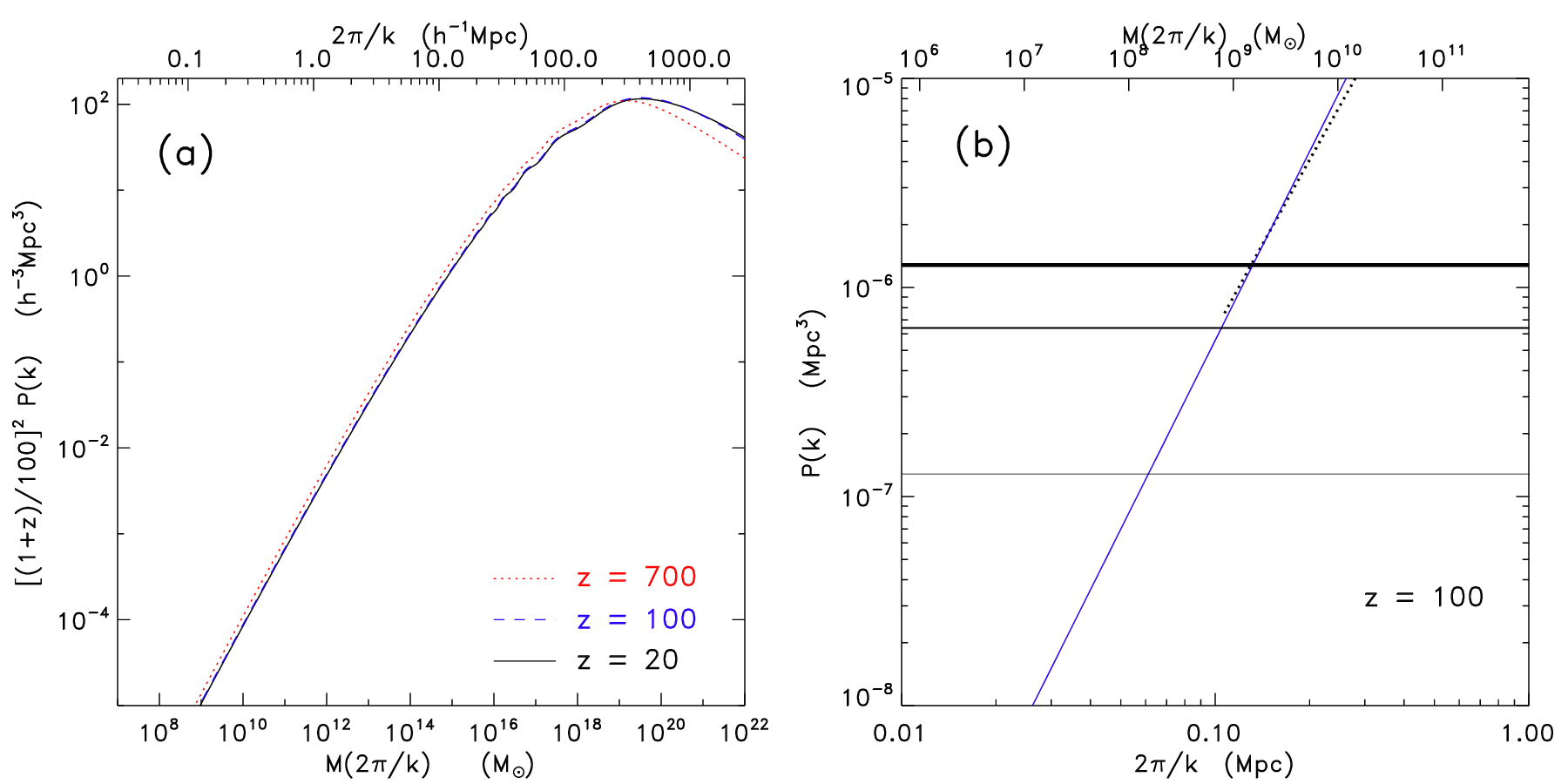}
\caption{\small
High-redshift structure formation in PBH cosmology.
{\it Top:} (a) Scaled power spectra for concordance (C)DM model at $z=700,100,20$, shown with red-dotted, blue-dashed and black-solid lines, respectively. (b) (C)DM power relevant for early small scale collapse, shown with black dotted line, compared to the fit according to Eq.~\ref{eq:power_small} (blue solid line). Thickest, thick and thin solid horizontal lines correspond to $f_{\rm PBH}M_{\rm PBH}=20, 10, 2\, M_\odot$, respectively. 
}
\label{fig:fig1ab}
\end{figure*}
If DM is at least partially made up of PBHs, these PBHs would generate an isocurvature component in the fluctuation power spectrum  \citep{Meszaros:1975,Meszaros:1980}. This granulation component is added to the inflationary fluctuation component of DM, $P_{\rm DM}$, which started with the Harrison-Zeldovich spectrum, later modified by post-inflationary evolution \citep{Afshordi:2003,Kashlinsky:2016}. The net power is given in \cite{Kashlinsky:2021}, which we write as:
\begin{equation}
P(k,z)= P_{\rm DM}(k,z) + 1.28\times 10^{-6}\left(\frac{ f_{\rm PBH}M_{\rm PBH} }{20M_\odot}\right)\left(\frac{1+z}{100}\right)^{-2} {\rm Mpc}^3 \mbox{\ .}
\label{eq:power}
\end{equation}
Here, $f_{\rm PBH}$ is the fraction of DM in PBHs and $M_{\rm PBH}$ is a suitably averaged PBH mass; specifically if PBHs are described by a mass function (unknown at present) $\zeta(m_\bullet)$, one has $M_{\rm PBH}=\int\zeta(m_\bullet)m_\bullet dm_\bullet/\int\zeta(m_\bullet)dm_\bullet$. 
The contribution of PBHs to the matter power spectrum is proportional to 
$f_{\rm PBH}M_{\rm PBH}$ so we will use
this factor to parametrize the effect of PBHs in the formation of early  halos.

Fig.~\ref{fig:fig1ab}a shows the (C)DM power spectra scaled by the Einstein-de~Sitter growth factor, starting at $z=700$ after matter-radiation equality. The figure shows that all power spectra, relevant for this discussion, collapse onto one template. At scales and times of importance for UHZ1, the small-scale power robustly follows a $\propto k^{-3}$ behavior, which is a consequence of the 3-dimensional nature of space coupled with the fact that matter density fluctuations on sub-horizon scales do not grow until the Universe has become radiation-dominated \citep{Meszaros:1974}.
At the small scales, or large $k$, relevant for this discussion, the DM power is shown in Fig.~\ref{fig:fig1ab}b, and the net power is analytically given by
\begin{equation}
P(k,z)=\left[2\times10^{-4}\left(\frac{\rm Mpc}{k}\right)^3\!+\!1.28\times 10^{-6}\left(\frac{f_{\rm PBH} M_{\rm PBH}}{20M_\odot}\right)\right]\left(\frac{1+z}{100}\right)^{-2} {\rm Mpc^3}\mbox{\ .}
\label{eq:power_small}
\end{equation}
The total mass inside a comoving radius of $2\pi/k=1$~Mpc is $M({\rm 1~Mpc})=1.7\times 10^{11}M_\odot$ for $h=0.71$, such that the evolution of regions with masses of $\lesssim 10^9M_\odot$ is robustly dominated by the extra power in the presence of PBHs.

Using the power in Eq.~\ref{eq:power_small}, we can compute the mean density contrast over a top-hat sphere with comoving radius $r_M$, containing mass $M$, as $\sigma_M^2(z) = \frac{1}{2\pi^2} \int_0^\infty P(k,z) W_{\rm TH}(kr_M) k^2dk$. Assuming a spherical collapse model, we estimate the masses $M$, where $\sigma_M(z) \geq \delta_{\rm col}=1.68$, such that these structures would turn around and collapse by redshift $z$, at the 1-$\sigma$ level. The ratio $\delta_{\rm col}/\sigma_M(z)$ is plotted vs. mass in Fig.~\ref{fig:fig1c} at $z=700$, for the standard CDM model and the various PBH-DM cases. The density field is shown at $z=700$, the time when radiation evolutionary effects have already ended, and the field is still well in the linear regime across the relevant scales, so its evolution in $z$ is self-similar with mass. From $z=700$ to $z=10$, as observed for UHZ1, the density field evolves as $\sigma_M\propto (1+z)^{-1}$.

\section{Formation of UHZ1-type Systems}
\label{sec:sec3}

Our proposal is that the formation of the observed central BH in UHZ1-type systems is based on the 2-body relaxation and mass segregation of the most massive PBHs in the early virialized halos. It only requires the mass function to be broad, so the more massive PBHs would be depleted from the halo by dynamical friction \citep{Chandrasekhar:1943} and will sink to the halo centers, producing a growing central BH irrespective of the surrounding baryonic matter that will eventually fragment into stars. Throughout the redshift range relevant to the discussion here the halo mass, $M_H$, extends over $10^6$ to $\gtrsim$10$^9\,M_\odot$. Broken power laws, log-normal distributions, and theoretically motivated spectra extensively considered in the literature \citep[e.g.][and references therein]{Carr:2021a} provide the current range of the mass spectra examples.

\begin{figure*}
\includegraphics[width=6.25in]{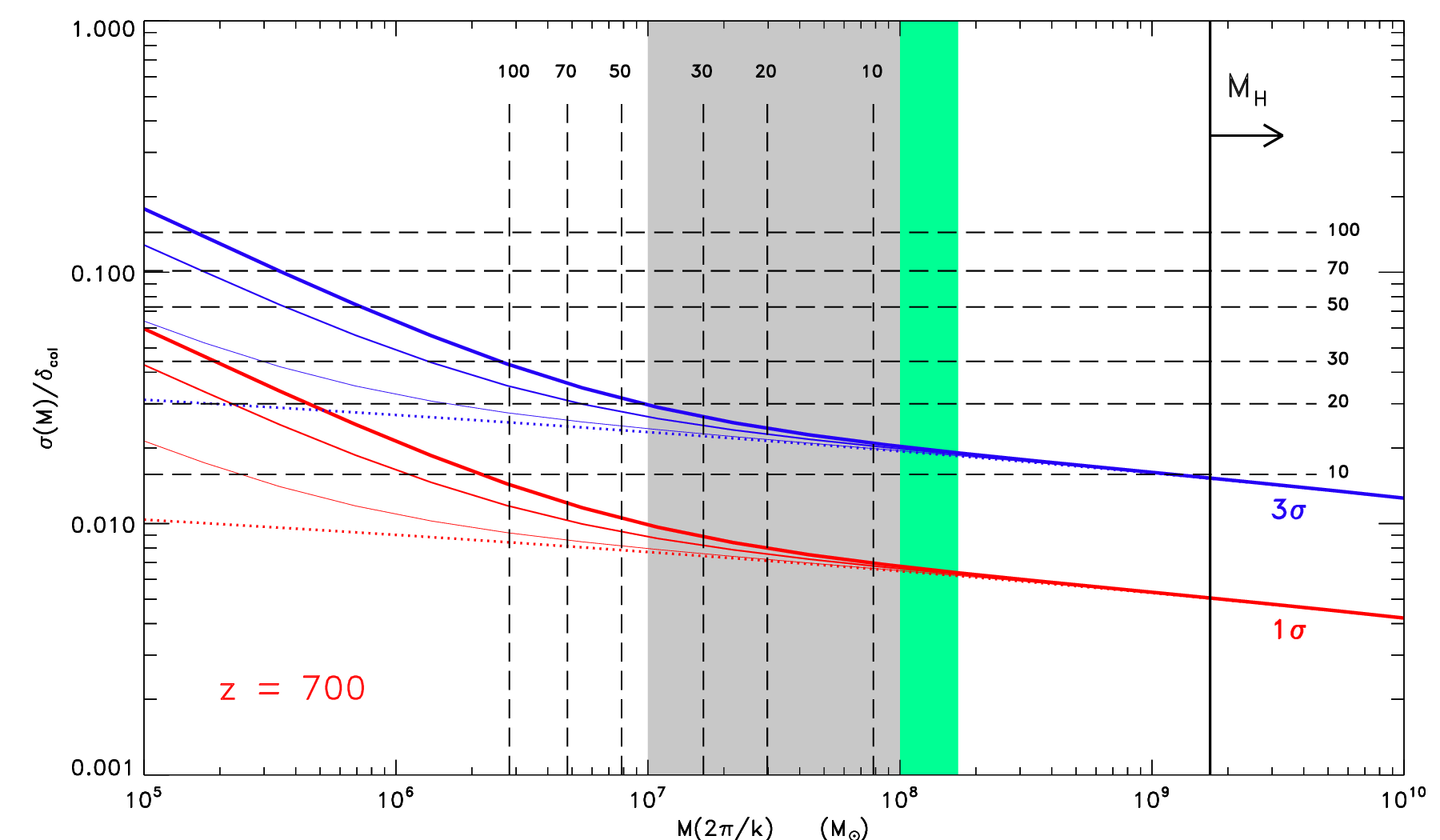}
\caption{\small
 Ratio of the ($1\sigma$) standard deviation of the density fluctuation amplitude to $\delta_{\rm col}=1.68$, shown in red at $z=700$, whereas the $3\sigma$ levels are shown in blue: dotted lines correspond to the power without PBH contribution, and thickest, thick and thin solid lines to PBH cosmologies with $f_{\rm PBH}M_{\rm PBH}=20, 10, 2\, M_\odot$, respectively.  Horizontal dashed lines indicate the density amplitude required at that redshift to collapse at the marked values of $z$. Vertical dashed lines show the mass with $T_{\rm vir}>10^4$\,K at the marked $z$, as required for the gas to cool, collapse and fragment into stars in the absence of H$_2$ \citep[see][]{Oh2002}. The estimated mass range of the UHZ1 BH is represented with the gray shading \citep{Bogdan:2024}, whereas the estimated mass range of the stellar component is marked in green \citep{Goulding:2023}. The vertical solid line marks the minimal halo mass implied by the stellar mass for UHZ1.
}
\label{fig:fig1c}
\end{figure*}

Fig. \ref{fig:fig1c} isolates the evolutionary regimes due to the granulation power where the halos: 1) collapse, virialize and start the gravitationally driven dynamical evolution of their PBH constituents  (where $\sigma_M(z)\geq \delta_{\rm col}$) and 2) form stars when the gas virial temperature enables gas collapse and fragmentation there. The dashed horizontal lines show the value of the ratio $\delta_{\rm col}/\sigma_M(z)$ required to collapse at the redshift marked at the end of each line. The PBHs inside such virialized halos can start reaching equipartition and mass segregation moving toward the center and forming one central SMBH \citep[see early relevant discussions in][]{Kashlinsky:1983}. Vertical dashes mark the masses that would have virial temperature $\gtrsim 10^4$\,K required to cool, collapse, and fragment into stars in the absence of H$_2$ formation and cooling. This threshold defines `atomic cooling halos', where collisional excitations of atomic hydrogen provide an efficient cooling channel even in the presence of radiation fields that would impact other coolants, such as H$_2$ \citep[e.g.,][]{Oh2002}. The gas inside such halos would then be able to form stars. The gray shaded area marks the mass-range of the UHZ1 black hole and the green shaded area shows the range inferred for its stellar content.

In the absence of the granulation power component arising from the LIGO-type PBHs as DM, there is little collapse and star formation until $z\lesssim 15-20$, whereas if the LIGO-type PBHs make up at least part of the dark matter, the first collapse of host halos can start as early as $z\sim 100$ in rare 
\textcolor{red}{$(\sim 4-5\sigma)$}, but sufficiently abundant regions of the density field, collapsing on mass scales of $\sim 10^6M_\odot$ at these epochs.
From Fig. \ref{fig:fig1c} we can read the fraction of halos that have collapsed and formed stars at different redshifts. For example, when the $f_{PBH}M_{PBH}=20$ thick solid blue line crosses the $z=20$ threshold for collapse (horizontal dashed line) for $M\leq 10^7M_\odot$, and indicates that $3\sigma$ density fluctuations at that mass scale would have collapsed by that redshift. Gas cooling followed by star formation would occur when the same blue line crosses the vertical dashed line that marks the same $z$. Gas could start forming stars by $z=20$ in halos of $M=3\times 10^7M_\odot$, with more rare density excursions: $3.3\sigma$ or larger
fluctuations in this example would be above the collapse threshold line, i.e., only $\sim 0.05\%$ of all halos on that halo mass scale would have collapsed.

Given the mass in stars, the implied total halo (virial) mass of UHZ1 must be at least $M_{\rm H}\gtrsim 1.7\times 10^9M_\odot$, marked with the vertical solid line, corresponding to $\gtrsim 3\sigma$ excursions of the underlying density field at $z=10$. 
Fig. \ref{fig:fig1c} shows that at $z=10$, on the mass scale associated with that of the UHZ1 system, there is little difference between the power spectra with or without the presence of PBHs. However the evolutionary differences at higher $z$ are very significant, depending on the PBH-DM contribution. As we shall discuss below, in collapsed halos, dynamical friction operating on PBHs in the high mass tail of their distribution could potentially produce the required central BH with a mass less than a few percent of the total mass of the halo by $z\sim10$, while this mechanism would not operate in $\Lambda$CDM. The following cosmogonical picture emerges from Fig.~\ref{fig:fig1c}: 1) the 
(lower mass) halos whose red and blue lines are above the horizontal dashed lines would be able to collapse at 1 and $3\sigma$ levels, respectively, and form virialized halos at the redshift marked next to the dashes. The gravitational evolution discussed below would lead to the formation of central massive BHs by $z\sim 10$, such as the one observed for UHZ1. 2) In the presence of only atomic hydrogen cooling, which requires $T\gtrsim10^4$\,K, stars would form only when halos can collapse, for the given power spectra parameters, to the right of the intersection points between the vertical dashed lines and the red and blue lines. This suggests a delayed onset of star formation in these halos, closer to $z\simeq10$. 
3) The points where the horizontal and vertical dashed lines of the same $z$ intersect determines the number of standard deviations in $\sigma_M$ of the objects that can collapse and start forming stars at the marked values of $z$. Fig.~\ref{fig:fig1c} further suggests that stars would form sufficiently ubiquitously only by $z\sim 10$ irrespectively of $f_{PBH}M_{PBH}$. What is relevant in PBH models is that halos collapse earlier than in $\Lambda$CDM
and since the mass distribution is broad, the most massive PBHs would fall to the center and build a SMBH.

We assume that the halos that reach the critical overdensity turn-around, collapse, and undergo violent relaxation \citep{Lynden-Bell:1967} to arrive at the Navarro-Frenk-White (NFW) density profile \citep{Navarro:1996}, but we also consider an isothermal density profile for completeness. In the presence of PBHs, the halo structure will be modified, but still remain close to the NFW profile at least for low $f_{\rm PBH}\lesssim 10^{-2}$, as has been established in cosmological N-body simulations \citep[][]{Zhang_PBH2024}. 
For halos with total mass $M_H(z)$ at redshift $z$, total radius $R_H$, and concentration parameter ${\cal C}\equiv R_H/r_{\rm core}$, the mean density is $\bar{\rho}=3M_H/(4\pi R_H^3)$, and their density profile can be expressed as $\rho_H(r)\equiv \bar{\rho} \Psi(r/R_H)$, where for
1) an isothermal halo \citep{Penston:1969,Gunn:1972}: $\Psi(x) =\frac{{\cal C}^3}{3A_{\rm ISO}}\frac{1}{1+{\cal C}^2x^2}$, with
$A_{\rm ISO}={\cal C}-\tan^{-1}{\cal C}$;\; and for
2) an NFW profile \citep{Navarro:1996,Navarro:1997}: $\Psi(x)= \frac{{\cal C}^3}{3A_{\rm NFW}}
\frac{1}{{\cal C}x(1+{\cal C}x)^2}$, with $A_{\rm NFW}=\ln(1+{\cal C})-{\cal C}/(1+{\cal C})$.
The circular velocity, $v_{\rm circ}(r)$, of the particles inside such halos is given by:
\begin{equation}
v_{\rm circ} ^2(r)= \frac{Gm_H(r)}{r} \equiv V_0^2 W(r/R_H)\mbox{ ,}
\label{eq:vofr}
\end{equation}
where $V_0\equiv 10 \sqrt{\Delta/200} H(z)R_H$. Furthermore, $m_H(r)$ is the halo mass within radius $r$, with a total mass of $M_H=m_H(R_H)$, and
\begin{equation}
  W(x) = \left\{\begin{array}{ll}
\frac{1}{A_{\rm ISO}}\times \left[\frac{{\cal C}x-\tan^{-1}({\cal C}x)}{x}\right] &    ~~~~~~~~~~~~~~~~  ({\rm isothermal})\\
& \\
\frac{1}{A_{\rm NFW}}\times \left[\frac{\ln(1+{\cal C}x)-{\cal C}x/(1+{\cal C}x)}{x}\right]  &   ~~~~~~~~~~~~~~~~ ({\rm NFW})\mbox{ ,} \\
\end{array}\right.
\label{eq:wofx}
\end{equation}
normalized to $W(1)=1$.
For isothermal halos the velocity dispersion and circular velocity remain constant outside the core, and follow $v_{\rm circ}(r)\propto r$ at $r<r_{\rm core}=R_H/{\cal C}$. For NFW profiles, the velocity increases as $v_{\rm circ}(r)\propto \sqrt{\ln ({\cal C}r)/r}$ outside the core, reaching a maximum at $r=2.2 r_{\rm core}$, while inside the core it decreases as  $v_{\rm circ}(r)\propto \sqrt{r}$. The NFW profile is empirically supported by observations of galaxy clusters with inferred concentration parameters ${\cal C}\simeq 10-15$ \citep{Atrio-Barandela:2008}. We note that the NFW profile (and its typical parameter values, e.g., ${\cal C}$) is numerically established for halos that form within the $\Lambda$CDM density field \citep[e.g.][]{Duffy:2008}; hence isothermal profiles are also presented.

The eﬀect of dynamical friction in the dynamics of BHs within dark matter halos is diﬃcult to
treat in numerical simulations. Due to the finite mass and force resolution the eﬀect of dynamical
friction is not resolved. Numerical eﬀects induce spurious motions on the dynamics of BHs and
correcting these artifacts requires repositioning of BHs in the minimum of the gravitational potential
\citep{Ragone-Figueroa:2018} or design prescriptions to account for the eﬀect
of dynamical friction \citep[see e.g.][]{Damiano:2024}. Simulations are also restricted to specific PBH masses and fractions and an analytical treatment is required to identify the key
physical processes. At the same time, the physics of dynamical friction is well understood \citep{Chandrasekhar:1943, Henon:1973} and analytical treatments of the individual orbital evolution under its influence have been well developed and are utilized here. This is the approach we follow.

For the NFW profile, the ratio of the mean halo density, $\bar{\rho}_H$, to the average matter density of the Universe at the time is $\Delta =\bar{\rho}_H(R_H)/\rho(z)$=200  \citep{Cole:1996}, so the halo radius is 
$R_H \simeq 0.15 (\Delta/200)^{-1/3} \left({M_H}/{10^6M_\odot}\right)^{1/3}[{(1+z)}/{21}]^{-1}$\,kpc. We adopt the same criterion for the isothermal distribution. The halo velocity dispersion is 
$V_0=\sqrt{\Delta/2} H(z)R_H\simeq 5.5 \sqrt{\Delta/200} \sqrt{{(1+z)}/{21}} \left({M_H}/{10^6M_\odot}\right)^{1/3}$ km~s$^{-1}$, with the halo
crossing time $t_{\rm cross}$=$R_H/V_0= H^{-1}(z)\sqrt{\Delta/2}$=$0.1 \sqrt{{\Delta}/{200}}\;t_{\rm cosm}(z)$. After virialization, the PBHs contained in such halos will start the 2-body relaxation leading to stellar dynamics evaporation and mass segregation due to dynamical friction. The gravitational stellar dynamical evaporation process in a system of $N_\bullet$ PBHs proceeds on a timescale of $t_{\rm ev}\simeq (N_\bullet/4\pi\ln N_\bullet) t_{\rm cross}$, 
much longer than it takes for most halos to get absorbed into a larger system at the next stage of gravitational clustering, occurring on the ($\sim 50$~Myr) free-fall time of the larger system, and is thus negligible for this discussion. However, if the PBHs do not have identical masses, the more massive ones with mass $m_\bullet\gg M_{\rm PBH}$ will undergo mass segregation on a timescale $< (M_{\rm PBH}/m_\bullet) t_{\rm ev}$, depending on the angular momentum of their orbit, and sink to the halo center. {\it Such halos will have at least about 10--30 crossing times at each $z$ to achieve mass segregation, enabling the assembly of suitably massive PBHs in their centers.}

While the PBHs with average mass, as given by their mass function established in the very early Universe ($z\gg10^5)$, will be undergoing slow relaxation via energy equipartition, leading (much later) to gravitational evaporation on time scales of $t_{\rm ev}>t_{\rm cosm}$, the more massive ones will transfer their kinetic  energy through dynamical friction to the less massive PBHs, and sink to the center.  The dynamical friction force \citep{Chandrasekhar:1943,Henon:1973}, acting on PBHs that are more massive than the average, will be in the (negative) velocity direction, $\hat{\mathbf{ v}}$, as given by:
\begin{equation}
\begin{array}{ll}
\mathbf{ F}&=- \hat{\mathbf{ v}} \frac{4\pi G^2(m_\bullet \ln \Lambda)\Delta \rho_{\rm cr}(z)}{v^2}  \Psi(r/R_H) f(<v)\\ 
&= - \hat{\mathbf{ v}} \frac{G(m_\bullet \ln \Lambda)}{R_H^2}\left(\frac{v}{V_0}\right)^{-2}f(<v)[3\Psi(r/R_H)] \mbox{\ .}\\
\end{array}
\label{eq:df}
\end{equation}
Here, $\ln \Lambda \sim \ln N_\bullet$ is the Coulomb logarithm, assumed to be constant \citep{Tremaine:1975,White:1976,Kashlinsky:1984}.
Equations of motion are given by eqs. 5, 7 of \cite{Kashlinsky:1984}, written here as
\begin{equation}
   \left\{\begin{array}{ll}
\ddot{r} &= -\frac{\dot{r}}{v} \frac{G(m_\bullet \ln \Lambda)}{R_H^2}\left(\frac{v}{V_0}\right)^{-2}[3\Psi(r/R_H)] f(<v)\\
&- \frac{V_0^2}{R_H}\left[\frac{W(r/R_H)}{r/R_H}\right]+ \frac{J^2}{r^3}\\
 & \\
\dot{J} &= -\frac{J}{v} \frac{G(m_\bullet \ln \Lambda)}{R_H^2}\left(\frac{v}{V_0}\right)^{-2}[3\Psi(r/R_H)] f(<v)\mbox{ ,} \\
  \end{array}\right.
 \label{eq:motion}
\end{equation}
where the total velocity is $v=\sqrt{\dot{r}^2 + J^2/r^2}$. We consider PBHs with apocenters at $r=R_H$; this encompasses most of the mass and gives a {\it lower} limit on the efficiency of the central BH formation. The equations of motion are then subject to the initial condition of $J(t=0)\equiv\lambda J_{\rm circ} = \lambda \sqrt{GM_HR_H}$, where $0\leq \lambda\leq 1$ is the starting orbital spin parameter, and initially varying between purely radial and purely circular orbits. Due to dynamical friction the orbits will eventually circularize as discussed in detail in \cite{Kashlinsky:1984}.

Inside each virialized halo we adopt for simplicity the distribution function for an isothermal sphere, ${\cal F}(E)\propto\exp[-E/(2V_0^2)]$, with $E$ being the total energy of each particle/PBH. However, several truncations are possible, which would lead to different evolutionary pathways, depending on the behavior of the velocity anisotropy as one approaches $R_H$. E.g., the King energy truncation \citep{King:1966}, ${\cal F}(E)\propto\exp[-E/(2V_0^2)]-1$, results in predominantly radial orbits near $R_H$, but a truncation in radius as in \cite{Kashlinsky:1988} would lead to predominantly circular orbits near $R_H$ and slower evolution. The distribution function of the overall halo will remain unchanged throughout the dynamical friction evolution of the few massive PBHs, whose individual energies will decrease under the action of the dynamical friction. Hence $f(E)$ is taken to be constant over time. Given that the PBHs of interest would have low $\lambda$ orbits, and that the most significant evolution driven by dynamical friction occurs at $r\ll R_H$, where $v\gg V_0$, to enable their sinking in less than a few cosmic times, we here do not explore further modifications from other possible truncations.

\begin{figure*}
\includegraphics[width=6.25in]{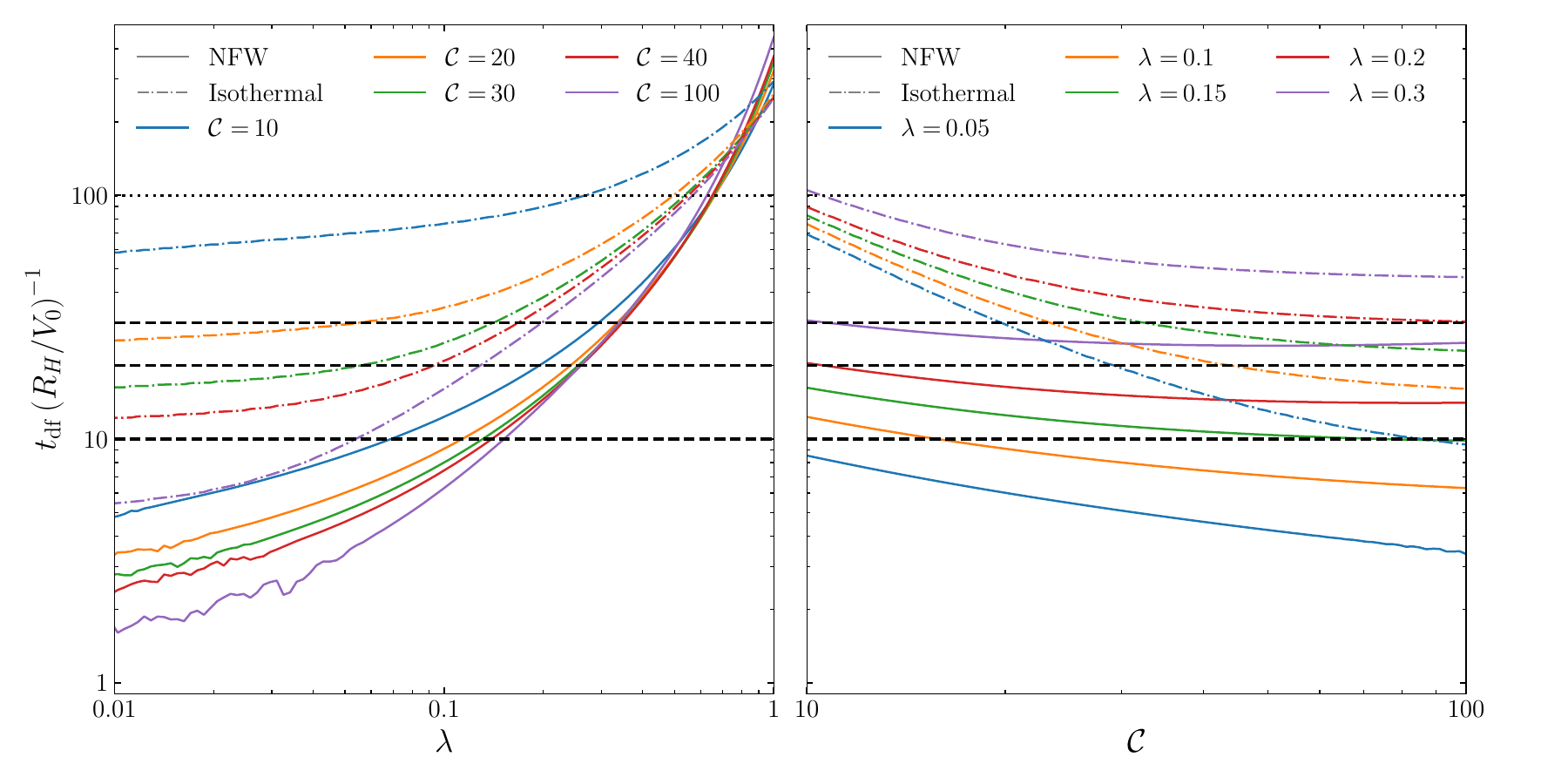}
\caption{\small
Dynamical friction collapse time $t_{\rm df}$ from solving numerically eqs. \ref{eq:motion}, in units of the crossing/orbital time, $R_{\rm H}/V_0$ as a function of $\lambda$ and ${\cal C}$. These results correspond to $m_\bullet \ln \Lambda/M_{\rm H}=0.01$ and can be rescaled to other values since $t_{\rm df}\propto (m_\bullet \ln \Lambda/M_{\rm H})^{-1}$. Dashed horizontal lines mark $(1,2,3)\times t_{\rm cosm}(z)$, whereas the dotted line indicates the time for an individual halo to remain intact from $z=50$ to $z=10$. {\it Left:} $t_{\rm df}$ vs. $\lambda$; NFW halo profiles with ${\cal C}=10,20,30,40, 100$, are shown with blue, orange, green, red, and purple solid lines, and isothermal halo profiles with the same concentrations are shown with corresponding dash-dotted lines. The isothermal, non-truncated distribution ${\cal F}(E)$ was adopted. {\it Right:} Same vertical axis vs. the concentration parameter, ${\cal C}$. Solid lines represent NFW profiles with $\lambda=0.05,0.1,.0.15,0.2,0.3$ in blue, orange, green, red, and purple, respectively, and dash-dotted lines correspond to isothermal profiles.
}
\label{fig:fig_df}
\end{figure*}

Fig. \ref{fig:fig_df} shows $t_{\rm df}$, the time required by dynamical friction to drive the BH of $m_\bullet \ln \Lambda/M_{\rm H}=0.01$ to the center, from integrating eq. \ref{eq:motion}, in the NFW and isothermal halos per eq.(\ref{eq:wofx}); $t_{\rm df}$ scales proportionally to $m_\bullet^{-1}$. Fig. \ref{fig:fig_df}, left panel, shows the collapse time vs. $\lambda$ for various concentration parameters ${\cal C}$ for NFW halos and an isothermal profile from \cite{Kashlinsky:1984}. The dashed horizontal lines mark $(1,2,3)\times t_{\rm cosm}(z)$ for the evolution to occur before the halo is incorporated in the next stage of gravitational clustering, when its structure would be modified. The horizontal dotted line specifically marks the time allowed for the halo to remain structurally undisturbed from $z=50$ to the observed $z=10.1$. In Fig.~\ref{fig:fig_df}, right panel, we show the same timescale vs. ${\cal C}$, for various $\lambda$. 
The scaling $t_{\rm df}\propto m_\bullet^{-1}$ allows to translate to the range of $m_\bullet$ that can collapse to the center for any PBH mass function. %, as e.g. presented in \cite{Byrnes:2018,Carr:2021}.

The mass accumulated in the center after $(1-3)\times t_{\rm cosm}$, before the halo is absorbed into the next stage of gravitational clustering, depends on: 1) the PBH mass function, 2) orbital spin parameter $\lambda$, 3) density profile, and 4) stellar distribution function ${\cal F}(E)$, where $E=\frac{1}{2} v_r^2 + \lambda^2 J_{\rm circ}^2/r^2 +\Phi(r)$, and the orbital angular momentum $J$ arising from the 2-dimensional tangential velocity component. Only the most massive PBHs will undergo dynamical friction, and the fraction of PBHs that sink to the center further depends on their angular momentum.
For a given distribution function ${\cal F}(E,J)$, the fraction of PBHs with spin parameter $\leq \lambda$ is $g(\leq \lambda) = \int_{v_r} dv_r \int_J {\cal F} dJ $. Adopting an isothermal distribution, ${\cal F}(E)=\frac{1}{\sqrt{2\pi}\sigma} \exp(-\frac{1}{2}v_r^2/\sigma^2 -J^2/R_H^2/\sigma^2)$, leads to $g(\leq \lambda)={\rm erf}(\lambda)$, which becomes $g\simeq \lambda$ at $\lambda<<1$.  For a purely isothermal distribution, objects faster than the escape velocity ($E>0)$ leave the system, rather than fall into the center. 
Given that Fig. \ref{fig:fig_df} shows that only the PBHs on orbits with $\lambda\ll 1$ are likely to sink to the center before the next stage of clustering, the approximation $g(\lambda)\sim \lambda$ is adequate for this discussion. 
The mass function of PBHs is unknown and is treated here as a freely adjustable function. 
An additional free parameter is the concentration ${\cal C}$ for the NFW profiles considered here. We can thus plausibly assume that a fraction $>0.2(\lambda/0.2)$ of all PBHs with $m_\bullet >10^{-4}M_{\rm H}(\ln\Lambda/10)$ would sink to the center in less than one Hubble time ($t_{\rm cosm}$), if halos have ${\cal C}\gtrsim20$; this fraction would rise to $0.5(\lambda/0.5)$, if $3\times t_{\rm cosm}$ were available for their individual evolution. In halos with ${\cal C}\sim 10$, this fraction becomes $\sim (0.5, 0.4)$ over (1,3)$\times t_{\rm cosm}$ for $m_\bullet >10^{-4}M_{\rm H}(\ln\Lambda/10)$. The timescales decrease towards  a larger PBH mass as $m_\bullet^{-1}$.

We can now tie this discussion to Fig.~\ref{fig:fig1c} which indicates the halo mass range that can collapse at any $z$, for various $f_{\rm PBH}M_{\rm PBH}$ and $(1,3)\sigma$ density excursions. 
   The figure shows that at the 3$\sigma$ level, halos of $M_{\rm H}\simeq (1.5\times 10^5, 3 \times 10^5, 7\times 10^5, 2.5\times 10^6, 10^7)M_\odot$, can collapse at $z=100, 70, 50,30,20$, if $f_{\rm PBH}M_{\rm PBH}=20M_\odot$, decreasing to $M_{\rm H}\simeq (\leq 10^5, 1.7 \times 10^5, 4\times 10^5, 1.3\times 10^6, 5\times 10^6)M_\odot$ at $z=100, 70, 50,30, 20$, if $f_{\rm PBH}M_{\rm PBH}=10M_\odot$. 
The Press-Schechter \citep{Press:1974} formalism suggests that a sufficient fraction of such halos would exist at $(3-5)\sigma$ levels to produce - for suitable PBH mass functions - at least a few central objects of $10^7-10^8M_\odot$ by $z=10$, as observed for UHZ1. E.g., at $3\sigma$ we can expect $\simeq 0.3\%$ of the density field to have the required parameters. 
Currently, theoretically-motivated PBH mass function proposals contain 10 to 20\% by mass in PBHs of $10M_\odot$ or higher \citep{Carr:2021}, i.e., those subject to dynamical friction, and 10-20\% of them will have the required low angular momentum, $\lambda\leq 0.1-0.2$. 
Dynamical friction would therefore drive $\sim 1-2\%$ of the total mass of the PBH component to the central 1~pc in $(0.5-1)\times t_{\rm cosm}$ after halo collapse. These clusters of PBHs at the centers of low mass halos will merge into larger units as hierarchical clustering proceeds. 
At $z\simeq 15$, halos of $10^9M_\odot$, corresponding to peaks of $3-4\sigma$, could have collapsed and given rise to concentrations of up to $\sim 10^7M_\odot$ in the central parsec by $z=10$, depending on the fraction, $f_{\rm PBH}$, of DM consisting of PBHs. 
Baryonic accretion may lead to a further increase in BH mass 
enhancing the efficiency of mass segregation further. 
An open question is how those PBHs will eventually merge to give rise to one central BH, since once at the center dynamical friction will no longer operate. The dynamics in a dense central cluster of PBHs may play a role here, similar to the stellar systems discussed in \cite{Begelman:1978,Begelman:1980}. 
More specifically, once the PBHs sink to near the center, the formation of one central, massive and growing BH could be similar to the various mechanisms discussed in \cite{Begelman:1978}, \cite{Kashlinsky:1983}. At present, there is no clear solution to how BHs overcome the last parsec problem and form tightly bound binary systems that can merge by gravitational wave emission, although we note that because of the dynamical friction evolution the PBHs in the central region will be on circular orbits where the gravitational wave emission is a steep function of their resultant orbital angular momentum $\propto j^7$. 
Furthermore, PBHs could be clustered and gravitational interaction could lead to the formation of BH pairs \citep{Chisholm:2011}. These PBH binaries will undoubtedly help to overcome the last parsec problem.

\begin{figure*}
\includegraphics[width=6.25in]{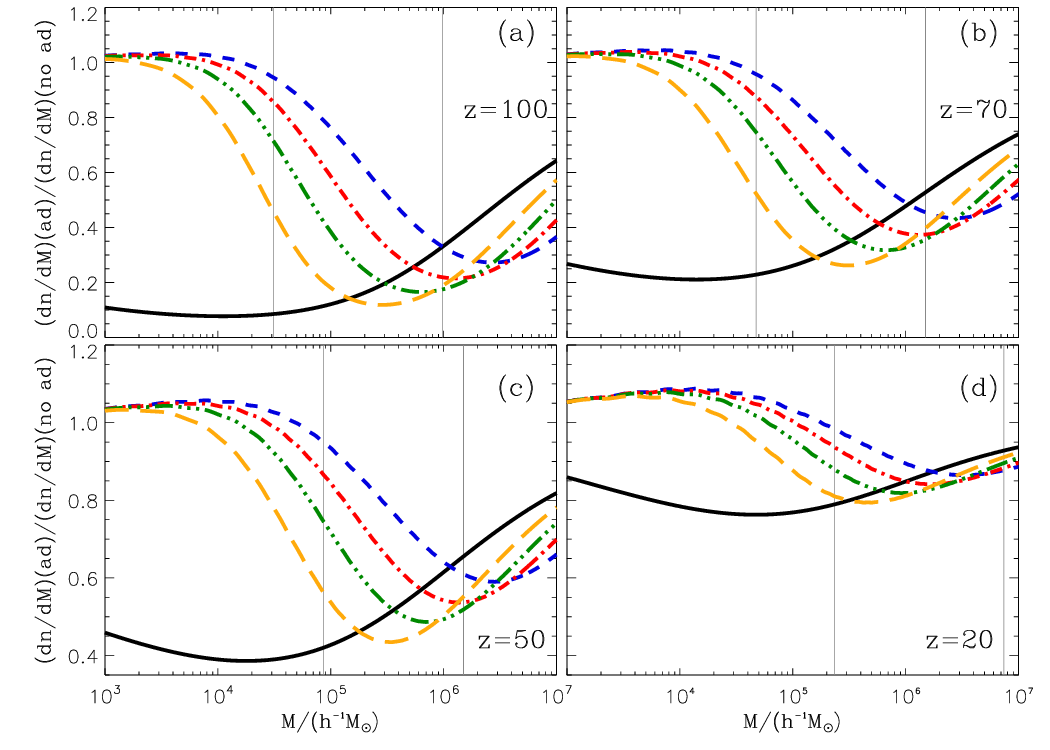}
\caption{\small
Impact of advection on the halo mass function. We compare the ratio of the number density of collapsed objects (per unit mass) for models with advection to models without it. Dashed (blue),
dot-dashed (red), triple dot-dashed (green), long dashed (gold) and solid (black)
lines correspond to $f_{\rm PBH}M_{\rm PBH}=(20,10,5,2,0)M_\odot$, respectively.
The vertical gray lines correspond to the masses of halos that have reached a
virial temperature $T_{\rm vir}=10^3$\,K (left), which allows H$_2$ formation and cooling, and $T_{\rm vir}=10^4$\,K (right), with atomic hydrogen cooling in the absence of H$_2$ \citep[see][]{Jaacks2019}.
The different plots correspond to different $z$, as indicated.}
\label{fig:fig3}
\end{figure*}
As shown in Fig.~\ref{fig:fig1c}, the halos of the relevant mass required to collapse at $z=10$ to explain the UHZ1 stellar contents would originate from $\gtrsim 3\sigma$ matter fluctuations within the DM power component, independently of the granulation power amplitude. The stellar component required to form at that epoch may be therefore independent of the PBH evolution, as discussed above, provided the baryonic gas can collapse and fragment into stars efficiently. The detailed processes of gas dynamics, star formation, and baryon-DM co-evolution are highly complex, and addressing this fully coupled co-evolution requires cosmological, hydrodynamical simulations \citep[see e.g.][]{Liu_PBH2022}. In general, simulations are always limited to the mass function and PBH fraction specific to each case and their results can not be extrapolated to other mass functions and PBH fractions. Our straightforward proposal highlights the physical processes underlying the evolution of halos containing a significant fraction of PBHs and the plausible mechanism determining star formation, so the results are easily extrapolated to different fractions of PBHs. 

We briefly discuss some plausible limits on the amount of stars formed. 
\color{black}{}
As pointed out by \cite{Tseliakhovich:2010}, for $\Lambda$CDM power spectrum the advection of small-scale perturbations by 
large-scale velocity flows at and near the baryonic Jeans scale may slow the infall of
baryons onto the DM potential wells, delaying the onset of star formation.  
These
2nd order evolutionary corrections are small if PBHs comprise at least a fraction of the DM, thus enabling efficient and rapid collapse \citep{Kashlinsky:2021,Atrio-Barandela:2022}.
To analyze the effect of advection, in Fig.~\ref{fig:fig3} we
compare the ratio of the number density of collapsed halos per
unit mass in the $\Lambda$CDM model with different
fractions of CDM and PBHs to the standard $\Lambda$CDM without advection.
The solid black line shows the effect of advection in the number
density of the standard model. The dashed lines correspond to models
containing PBHs as well. For larger fractions of PBH and larger masses of BH,
the number density of less-massive halos is consistently higher than in $\Lambda$CDM
at all $z$, where the difference is more significant at higher redshifts. 
Since less massive halos are formed more abundantly than in the concordance model,
if all other physical conditions are kept equal, the presence of PBHs
would result in an earlier collapse of the low-mass halos and an earlier baryon
infall, thus favoring the formation of the first stars.  

We therefore assume that no delay in the subsequent star formation occurs due to such DM-baryon streaming effects, forming sufficient stars by the target redshift ($z\sim 10$) inside UHZ1-type host halos.
Fig. \ref{fig:fig1c} shows that collapse and star formation at $z\sim 10$, observed for UHZ1, is driven by the inflationary DM power component and not by the granulation term. The implied halo mass required to contain the observed baryonic mass in stars is then $M_{\rm H}>10^9\epsilon_*^{-1} M_\odot$, where $\epsilon_*$ is the star formation efficiency. Whereas for the lower limit on host halo mass, $M_{\rm H}\sim 2\times 10^9M_\odot$, the stars would have to form in this 3-$\sigma$ halo at their maximal efficiency close to $\epsilon_*\sim 1$, the matter density spectrum is sufficiently shallow with $M_{\rm H}$ in this regime. This would imply that by considering a slightly more biased density region, corresponding to about 3.5-$\sigma$, the required efficiency would decrease to $\epsilon_*=0.1$. Such systems, while rare, would still be sufficiently abundant at the observed $z\sim 10$.

\section{Discussion}

If LIGO type PBHs make up at least part of the DM, they inevitably generate a granulation component that dominates the power produced during inflation, which decreases as $\propto k^{-3}$ on the scales relevant for early halo collapse. After discussing the modifications this component generates in the formation of collapsed objects at $z\gg 10$, we presented a framework where dynamical friction, operating in a virialized halo containing PBHs with different masses, inevitably drives the most massive ones into its center. The efficiency of this mass segregation depends on currently unknown parameters, primarily the PBH mass spectrum, and the detailed density distribution within the first collapsed halos containing the PBHs. We showed that for certain plausible parameters the evolution leads to a dominant massive BH in the halo center already at $z\gg10$. 
As more
PBHs reach the central region, they will cluster and may form a central massive BH. During hierarchical clustering, 
less massive halos merge into bigger units and, subsequently, dynamical friction drives the BH 
clusters to the center of the final system. The deceleration produced by dynamical friction is most effective within or near the denser halo center since the number of lower-mass objects and their velocities are higher. 
A mass concentration $M_{\rm BH}\sim 10^{7-8}M_\odot$ at the center  in $\sim 450$~Myr can be achieved without requiring super-Eddington accretion rates. 
One can plausibly argue that the mass ratio of the central BH to the mass in stars will be in the range $\sim 0.01-0.1$, as observed for UHZ1. 

Gravitational microlensing observations have been conducted to constrain the abundance of PBHs in Galactic halo.
The measured optical depth towards the Magellanic Clouds (MCs) were claimed to appear compatible with the few measured
events being due to stars  \citep{Mroz:2024,Mroz:2024a,Mroz:2025} because 
the OGLE observations, using only one photometric band, failed to detect long-duration events expected due to massive BHs in the halo of the Galaxy. The assumptions behind these results and other Galactic microlensing models have been addressed in \cite{Hawkins:2020,Garcia-Bellido:2024} showing explicitly that the microlensing constraints do not 
rule out the PBHs here \citep{Hawkins:2025}. Additionally, in our model, dynamical friction would have depleted the Galaxy halo of its most massive
BHs and those in the MCs would have been driven to the central region where the crowded field would compromise the 
photometry and these events could have been missed. High-cadence observations did
find an excess of short lensing events towards the bulge but not towards the MCs,
making it difficult to explain the results with either low mass PBHs or free floating 
planets. Also, since PBHs could be in pairs \citep{Chisholm:2011}, their lensing effect would be missed when searching for single-lens events. 
Mergers of black hole pairs would lead to the formation of SMBHs much earlier than in the low-mass and high-mass seed scenarios in the standard $\Lambda$CDM model. 
Spectral distortions of the CMB from accreting PBHs are too small to be detected by COBE/FIRAS as well as future instruments \citep{Ali-Haimoud:2017a}. The overall prospects are discussed in \cite{Kashlinsky:2019b}.

Given the uncertainties, we do not expect a
large number of UHZ1-type objects to form. But dynamical friction is  inevitable and operates as the PBH halos collapse and virialize. Therefore, one would expect the number of these objects to grow with time since 
halos in lower-density regions will also collapse. The recently uncovered population of numerous faint, broad-line AGNs at $z>5$ \citep{Onoue:2023,Kocevski:2023} with luminosities 2-3 orders of magnitude below those of bright quasars at similar redshifts and powered by BHs with masses $10^{6-8}M_\odot$, naturally 
fits into this framework. These Little Red Dots (LRDs) are characterized
by a compact morphology, a peculiar shape of their spectral energy distribution, and
$M_{\rm BH}/M_\star\sim 0.1-0.01$. Their
cosmic abundance is orders of magnitude higher than that of bright quasars, but their number
appears to drop drastically at $z<4$  
\citep{Kocevski:2024,Taylor2024}. Although UHZ1-type objects would require $3-5\sigma$ density peaks to
form by $z\sim 10$, LRDs of similar BH mass and BH-to-stellar mass ratio
would only require $2-3\sigma$ thresholds to have collapsed by redshifts $z\sim 4-8$, as shown in Fig.\ref{fig:fig1c}. In this context, both UHZ1 and LRD objects could be formed by the same mechanism.

\section*{Acknowledgements}

A.K. was supported by NASA under award 80GSFC24M0006 and acknowledges NASA/12-EUCLID11-0003 ``LIBRAE: Looking at Infrared Background Radiation Anisotropies with Euclid". F.A-B. acknowledges grants PID2024-158938NB-I00 by AERDF ``A way of making Europe" and by MCIN/AEI/10.13039/501100011033, as well as SA083P17 by the Junta de Castilla y Le\'on. D.M-G. was supported by a grant from the Fondo Social Europeo Plus, Programa Operativo de Castilla y Le\'on and the Consejer{\'\i}a de Educaci\'on, Junta de Castilla y Le\'on.

\section*{Data Availability} Data available on request.

\bibliographystyle{aasjournal}
		\bibliography{refs}



\end{document}